\documentclass[twocolumn,prb]{revtex4}
\usepackage{amsfonts}
\usepackage[T1]{fontenc}
\usepackage{amsmath,amsbsy,amssymb,graphicx}
\usepackage{times}
\let\mathbf=\boldsymbol
\def\section#1{\bigskip\leftline{\textbf{#1}}}

\begin{document}

\title{{\Large Quantized Anomalous Hall Effects in Skyrmion Crystal\\}}
\author{Keita Hamamoto$^1$, Motohiko Ezawa$^1$ and Naoto Nagaosa$^{1,2}$}
\affiliation{$^1$Department of Applied Physics, University of Tokyo, Hongo 7-3-1, 113-8656, Japan}
\affiliation{$^2$Center for Emergent Matter Science (CEMS), ASI, RIKEN, 
Wako 351-0198,Japan }

\begin{abstract}
We theoretically study the quantized anomalous Hall effect (QAHE) in skyrmion crystal (SkX) 
without external magnetic field. 
The emergent magnetic field in SkX  could be gigantic as much as $\sim4000$T
when its lattice constant is $\sim1$nm. 
The band structure is not flat but has a finite gap in the low electron-density regime. 
We also study the conditions to realize the QAHE for the skyrmion size, carrier density, 
disorder strength and temperature.
Comparing the SkX and the system under the corresponding uniform magnetic field, 
the former is more fragile against the temperature compared with the latter since 
the gap is reduced by a factor of $\sim$ 1/5, while they are almost equally robust against 
the disorder. Therefore, it is expected that the QAHE of SkX system is realized 
even with strong 
disorder at room temperature when the electron density of the order of one per a skyrmion.
\end{abstract}

\maketitle

Magnetic skyrmion is a topological spin texture in ferromagnets\cite{SkRev}. 
After the early theoretical proposals in magnets~\cite{Bogdanov,BogdanovB,BogdanovC}, 
the study of magnetic skyrmion is growing rapidly since it was discovered 
experimentally~\cite{Mol,Yu,Heinze}.
The periodic array of skyrmions, i.e., a skyrmion crystal (SkX), is realized at
interfaces\cite{Heinze} or in bulk chiral magnets such as B20 compounds\cite{Mol,Yu}.
An emergent magnetic field, generated in the background skyrmion spin texture.
Namely, a skyrmion has one magnetic flux $\Phi_0=h/e$ acting on the conduction electrons 
coupled to it. When the skyrmions form a periodic lattice, i.e., a SkX, 
the emergent magnetic field reaches $\sim 4000$T 
{\it assuming} the uniform averaged flux 
for the skyrmion size and the lattice constant of SkX of the order of $\sim1$nm. 
The effective magnetic field is proportional to $\lambda^{-2}$, where $\lambda$ 
is the skyrmion radius.  
Since the size of the skyrmion is 3nm for MnGe~\cite{Kanazawa}, 
18nm for MnSi~\cite{Lee},  and 70nm for FeGe~\cite{FeGe}, the corresponding
emergent magnetic field is $\sim 1100$T, $28$T, and $1$T, respectively. 

This emergent magnetic field   
leads to the Hall effect\cite{Shulz}.
Most of the studies focus on the Hall effect in metallic systems 
with large electron density\cite{Lee,Li,Kanazawa,Neu}. 
This so-called topological Hall conductivity $\sigma_{xy}$
is usually small compared with the longitudinal conductivity $\sigma_{xx}$, i.e., 
the Hall angle $\sigma_{xy}/\sigma_{xx}$
is typically of the order of $10^{-2}$ at most\cite{Kanazawa}. 

Up to now, we regard skyrmions as the source of the real space emergent 
magnetic field. When the size of the skyrmion becomes comparable 
to the mean free path, it is expected that the crossover from the real to momentum space 
Berry curvature occurs. 
One can regard the case of pyrochlore ferromagnet as the limit of 
large emergent magnetic field, where the spin chirality is defined for each 
unit cell of tetrahedron~\cite{Taguchi}.
In this case, there is no Landau Level (LL) formation, but the
band structure is formed by taking into account the solid angle of the spin, 
and the intrinsic anomalous Hall effect appears whose conductance is given by the integral of 
the Berry curvature in momentum space\cite{Onoda}. 
A more drastic example is the quantized anomalous Hall 
effect (QAHE) in magnetic topological insulator (TI), where the surface state
with gap opening due to the exchange coupling to the magnetic ions produces
the quantized Hall conductance of $e^2/h$ without the external magnetic 
field\cite{QAHE}. 

It is expected that the SkX offers an ideal laboratory to study 
QAHE from a unified viewpoint since one can change the
size of the skyrmion, the mean free path, and even the carrier concentration 
by gating at the interface, to reveal the crossover between real and momentum 
Berry curvature and stability of the quantized Hall conductance 
as these conditions are changed. 

In this paper we theoretically explore the emergence of the QAHE in the SkX 
without external magnetic field. 
Landau Levels are not formed in the present system
since the emergent magnetic field is not uniform.
Nevertheless, the band structure contains several well separated bands 
in the low electron-density regime, 
where each band has a Chern number $C=-1$. 
Consequently the emergence of the QAHE is predicted.
We point out that the lowest and next-lowest bands are well described by the Dirac theory.
On the other hand, Hall plateaux disappear in the large electron-density regime because 
of the overlap of bands.
We also clarify the conditions of QAHE for 
the skyrmion size, carrier density, disorder strength and temperature.

\textbf{The model:}
We start with a free electron system coupled with the background spin texture 
$\mathbf{n}_i$ by Hund's coupling\cite{Anderson,Ohgushi,Onoda}.
The Hamiltonian is given by the double-exchange model,
\begin{equation}
H=\sum_{ij}t^{ij}c^{\dagger}_ic_j -J \sum \mathbf{n}_ic^{\dagger}_i\mathbf{\sigma}c_i,
\end{equation}
where $c^{\dagger}_i$ ($c_i$) is the two-component (spin up and spin down) 
creation (annihilation) operator at the $i$ site, $t^{ij}$ is the transfer integral between
nearest-neighbor sites, $J$ is the Hund's coupling strength between the electron spin 
and background spin texture and $\mathbf{\sigma}$ denotes the Pauli matrix.
When Hund's coupling is strong enough $J \gg t^{ij}$, the spin of the hopping electron is
forced to align parallel to the spin texture\cite{Ohgushi,Onoda}. 
The wave-function $|\chi (\mathbf{r})\rangle$ of the conduction electron at 
$\mathbf{r}$ corresponding to the localized spin $\mathbf{n}(\mathbf{r})$ is given by
\begin{equation}
|\chi (\mathbf{r})\rangle =(\cos\frac{\theta (\mathbf{r})}{2},
e^{i\phi (\mathbf{r})}\sin\frac{\theta (\mathbf{r})}{2})^t,
\end{equation}
where we have introduced the polar coordinate of the spin configuration 
$\mathbf{n} =(\cos\phi\sin\theta,\sin\phi\sin\theta,\cos\theta )$.
Then the effective transfer integral is obtained, 
\begin{equation}
t^{ij}_{\text{eff}}=t\langle\chi_i|\chi_j\rangle =te^{ia_{ij}}\cos\frac{\theta_{ij}}{2},
\end{equation}
where 
$a_{ij}=(\phi_i-\phi_j) (1-\cos\frac{\theta_i+\theta_j}{2} )/2$ 
is the vector potential generated by the spin between the $i$ and $j$ sites,  
and 
$\theta_{ij}$ is the angle between the two spins. 
The effective tight-binding Hamiltonian is obtained as
\begin{equation}
H=\sum_{ij}t^{ij}_{\text{eff}}d^{\dagger}_id_j .
\label{Hamil}
\end{equation}
where $d^{\dagger}_i$ ($d_i$) is the spinless creation (annihilation) operator at the $i$ site.
One can easily see
that the spin chirality is absent for collinear and coplanar
spin alignment.

\begin{figure}[!t]
\centerline{\includegraphics[width=0.5\textwidth]{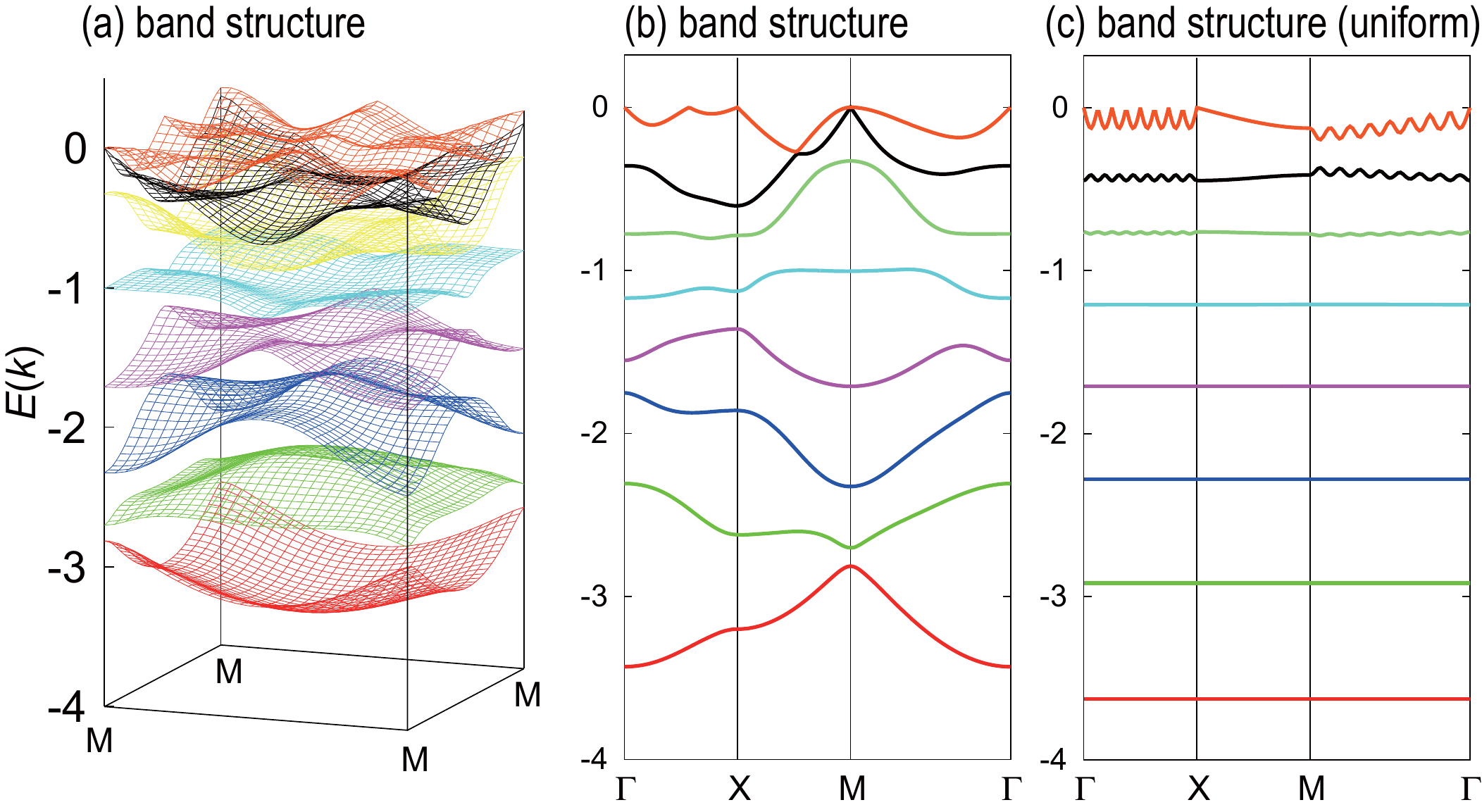}}
\caption{The band structure in the presence of SkX. We have set $\lambda =2$. The horizontal axes are momentum $k_x$ and $k_y$, while the vertical axis is the energy. There are $4\lambda^2 =16$ bands. We show (a) a bird's eye view and (b) a cross section of the lowest $8$ bands. (c) The band structure of a tight-binding model with the uniform magnetic flux, which forms the almost flat Landau levels except near the center of the energy 0.} 
\label{FigSkXBand}
\end{figure}

\textbf{Skyrmion crystal:}
We consider a background spin texture $\mathbf{n}(\mathbf{r})$ made of a square SkX. 
Each skyrmion has a nontrivial topological number\cite{Raja}.
The skyrmion profile is well assumed as $\theta (\mathbf{r})=\pi (1-r/\lambda )$ 
for $r<\lambda$ and  $\theta (\mathbf{r})=0$ for $r>\lambda$. 
The emergent magnetic field is produced by the spin texture 
since it has a finite solid angle\cite{SkRev,Volovik,Zang,Zhang}, 
\begin{equation}
b_z(\mathbf{r})=\frac{\hbar}{2e}\mathbf{n}\cdot (\partial_x \mathbf{n}\times\partial_y \mathbf{n}) =\frac{\hbar}{2e}\frac{\pi}{r\lambda}\sin\pi (1-r/\lambda )
\end{equation}
for $r<\lambda$ and $b_z=0$ for $r>\lambda$. 
It does not depend on the azimuthal angle $\phi$.
The total magnetic flux is
\begin{equation}
\int_0^\lambda d^2\mathbf{r} b_z(\mathbf{r})=\Phi_0,
\end{equation}
with $\Phi_0=h/e$, and is independent of the skyrmion radius $\lambda$.

\textbf{Band structure:}
The band structure can be obtained by numerically diagonalizing the Hamiltonian (\ref{Hamil}) 
in the unit cell with the size $2\lambda\times 2\lambda$ in the presence of the SkX. 
The Brillouin zone is given by $-\pi/2\lambda <k_x<\pi/2\lambda$ and  
$-\pi/2\lambda <k_y<\pi/2\lambda$.
Here we take the unit with the lattice constant $a=1$.
Then the unit cell has dimensionless area $(2\lambda)^2$.  
We show the band structure in the presence of SkX in Figs.\ref{FigSkXBand}(a) and (b), 
which are warped and the number of bands is $4\lambda^2$. 
There are finite gaps between two successive bands for lower bands.
However, the band overlap starts at a higher band.
In Fig\ref{FigSkXBand}(c) shown the band structure 
of the corresponding uniform mean magnetic field.
Here, since one skyrmion exists per area  $4\lambda ^2$,
the mean magnetic field $\bar{b}$ is given by $\Phi_0/4\lambda^2$.
It is seen that there is almost no energy dispersion, i.e.,
the Landau level formation, with the energy separation 
$\Delta_0$ which slightly depends on the Landau level index. 

\begin{figure}[!t]
\centerline{\includegraphics[width=0.5\textwidth]{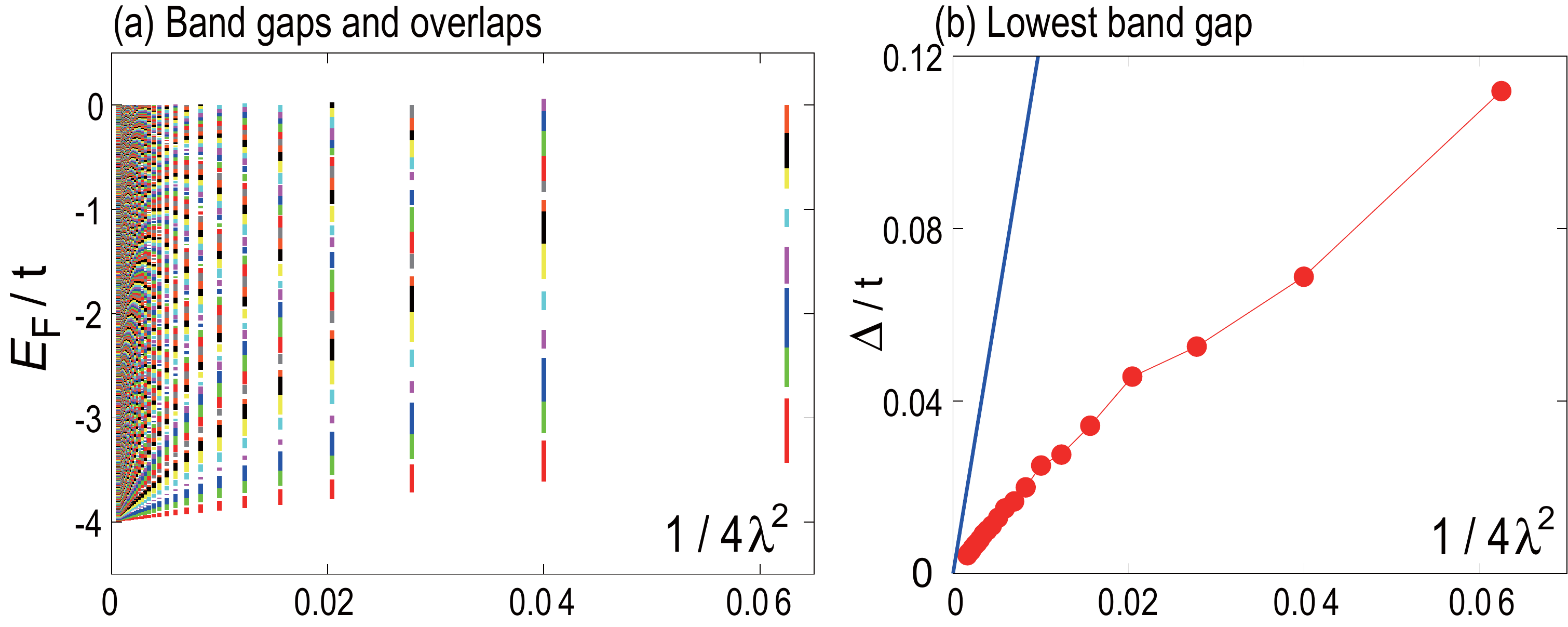}}
\caption{(a) Band gaps and overlaps for various sizes $\lambda$ of the skyrmions.  
(b) The $(2\lambda)^{-2}$ dependence of the lowest band gap $\Delta$. 
The gap is proportional to $(2\lambda)^{-2}$. The band gap of the system
with the corresponding uniform mean magnetic flux is shown in the blue line.}
\label{FigGap}
\end{figure}

We show the gaps and overlaps of bands in Fig.\ref{FigGap}(a) for 
various skyrmion radius  $\lambda$, where the bands are marked in color bar 
and the band gaps are denoted by the white blanks.
We find that the band gap shows a scaling behavior for the skyrmion radius.
The band gap between the lowest and second-lowest bands always opens.

We show the $\lambda$ dependence of the lowest band gap $\Delta$ in Fig.\ref{FigGap}(b). 
It is proportional to $1/\lambda^2$. 
There is some deviation from the linear fit in small $\lambda$, 
which is probably due to the finite size effect.
We can compare this band gap with the Landau level separation $\Delta_0$ 
shown in Fig.1(c).  
The lowest Landau level gap $\Delta_0$ is given by $0.71$t for $\lambda=2$, 
while the lowest gap of the QAHE in SkX is given by $\Delta=0.11$t.
The linear relation in the small $1/4\lambda^2$ region of Fig. \ref{FigGap}(b) 
indicates the relation $\Delta \cong \Delta_0/5$. 

\begin{figure}[!t]
\centerline{\includegraphics[width=0.5\textwidth]{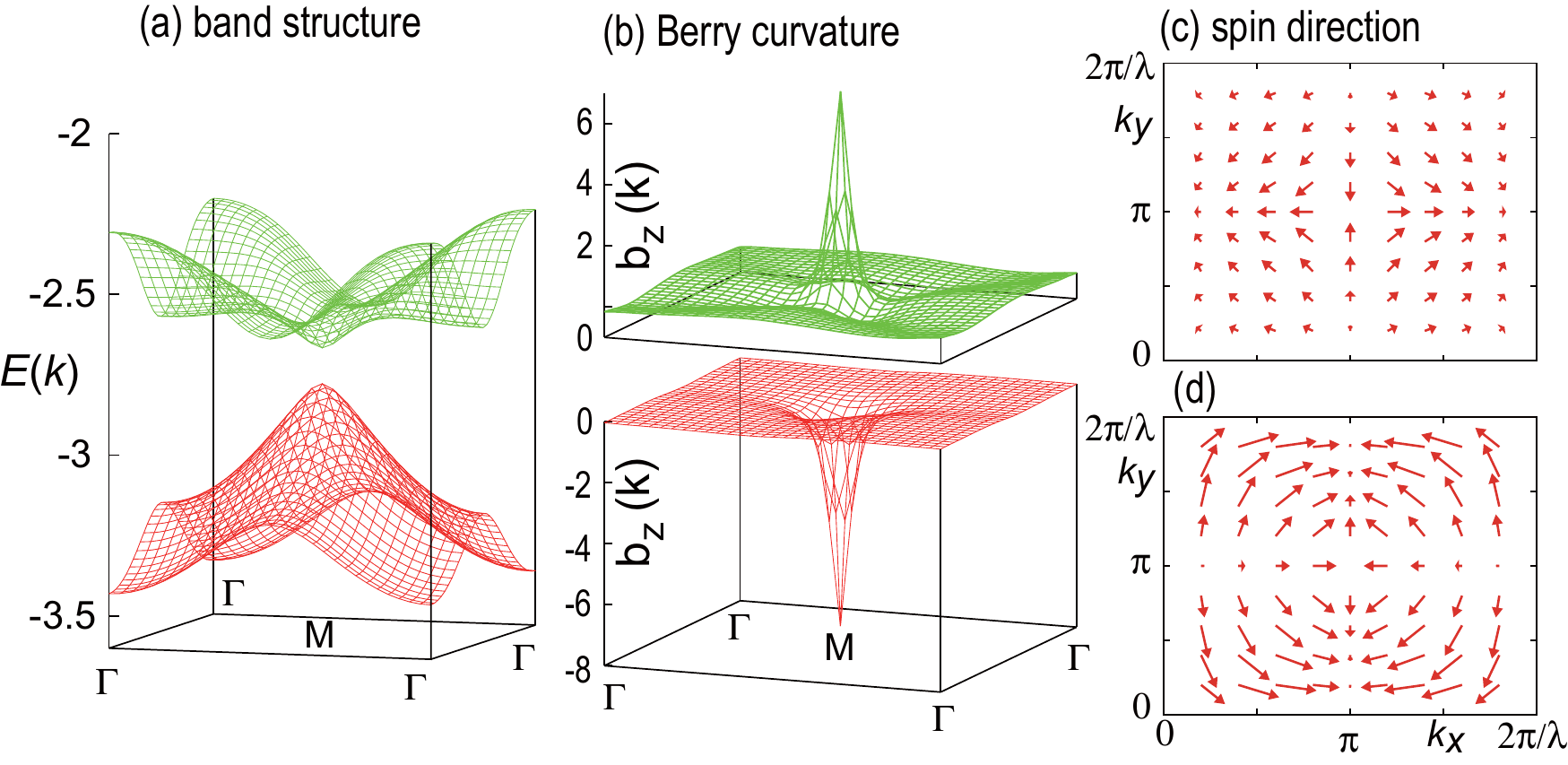}}
\caption{(a)  Band structure of the lowest and second-lowest bands in the vicinity of the $M$ point. It has the Dirac-cone shape.
(b) The momentum distribution of the Berry curvature in the lowest and second-lowest bands. The Berry curvature takes the largest value at the $M$ point, where the Chern numbers $C=- 1/2$ and $+1/2$ are generated for these bands, respectively. 
The residual Chern numbers $C=-1/2$ and $-3/2$ arise away from the $M$ point. The spin directions of the band structure of (c) the conduction and (d) valence bands.
They clearly exhibit the anti-vortex structure at the $M$ point.}
\label{FigBerry}
\end{figure}

\textbf{Berry curvature:}
We focus on the lowest and second-lowest bands [Fig.\ref{FigBerry}(a)]. 
We show the spin direction determined by 
$\langle \psi(\mathbf{k})|\sigma|\psi(\mathbf{k})  \rangle$ 
for conduction and valence bands in Fig.\ref{FigBerry}(c) and (d), 
respectively, where the anti-vortex structure is 
evident at the $M$ point. 
It implies that the spin texture has a nontrivial Berry curvature.

We may define a "gauge potential" in the momentum space,
$a_{k}(\mathbf{k}) =-i\left\langle \psi (\mathbf{k})
\right\vert \partial _{k}\left\vert \psi (\mathbf{k})
\right\rangle$,
for Bloch state $\left\vert \psi (\mathbf{k})\right\rangle $, 
which is properly called the Berry connection. Then we may define the
"magnetic field" or the Berry curvature by
$b_z(\mathbf{k}) =\partial_{k_x}a_{y}(\mathbf{k}) -\partial_{k_y}a_{x}(\mathbf{k})$.
The Chern number is the integral of the Berry curvature over the first Brillouin zone, 
$\mathcal{C}=\frac{1}{2\pi }\int d^{2}k b_z(\mathbf{k})$.

We show the momentum dependence of the Berry curvature in Fig.\ref{FigBerry}. 
It takes a large value in the vicinity of the $M$ point, which is 
$(k_x,k_y)=(\pi/2\lambda,\pi/2\lambda)$. The sign of the Berry curvature is negative 
for the lowest band, while it is positive for the second-lowest band. 
Accordingly, the sign of the Berry curvatures are opposite between the lowest and 
second-lowest bands.

Each band has one unit Chern number $C=-1$. This fact can be understood in terms of 
adiabatic connection from the Landau level. In our system, the emergent magnetic 
field has a space dependence. We consider an adiabatic pass from the uniform magnetic 
field to the space-dependent magnetic field induced by SkX. This deformation has no 
singularity. Thus the band structure in Fig.1 can be obtained from the deformation 
of the Landau level. Each Landau level has the same Chern number. 
Accordingly, each band of our system also has the same Chern number, which is $C=-1$.

The structure of the lowest and second-lowest bands has precisely the shape of the 
Dirac cone in the vicinity of the 
$M$ point $(k_x,k_y)=(\pi/2\lambda,\pi/2\lambda )$ 
as in Fig.\ref{FigBerry}(a). 
This suggests that electrons are described by the Dirac theory
$H=\hbar v(-k_x\sigma_x+k_y\sigma_y)+m\sigma_z, \label{DiracA}$
where the mass $m$ is related to the gap $\Delta$ between the lowest 
and second-lowest bands by $\Delta=2|m|$.

\begin{figure}[!t]
\centerline{\includegraphics[width=0.5\textwidth]{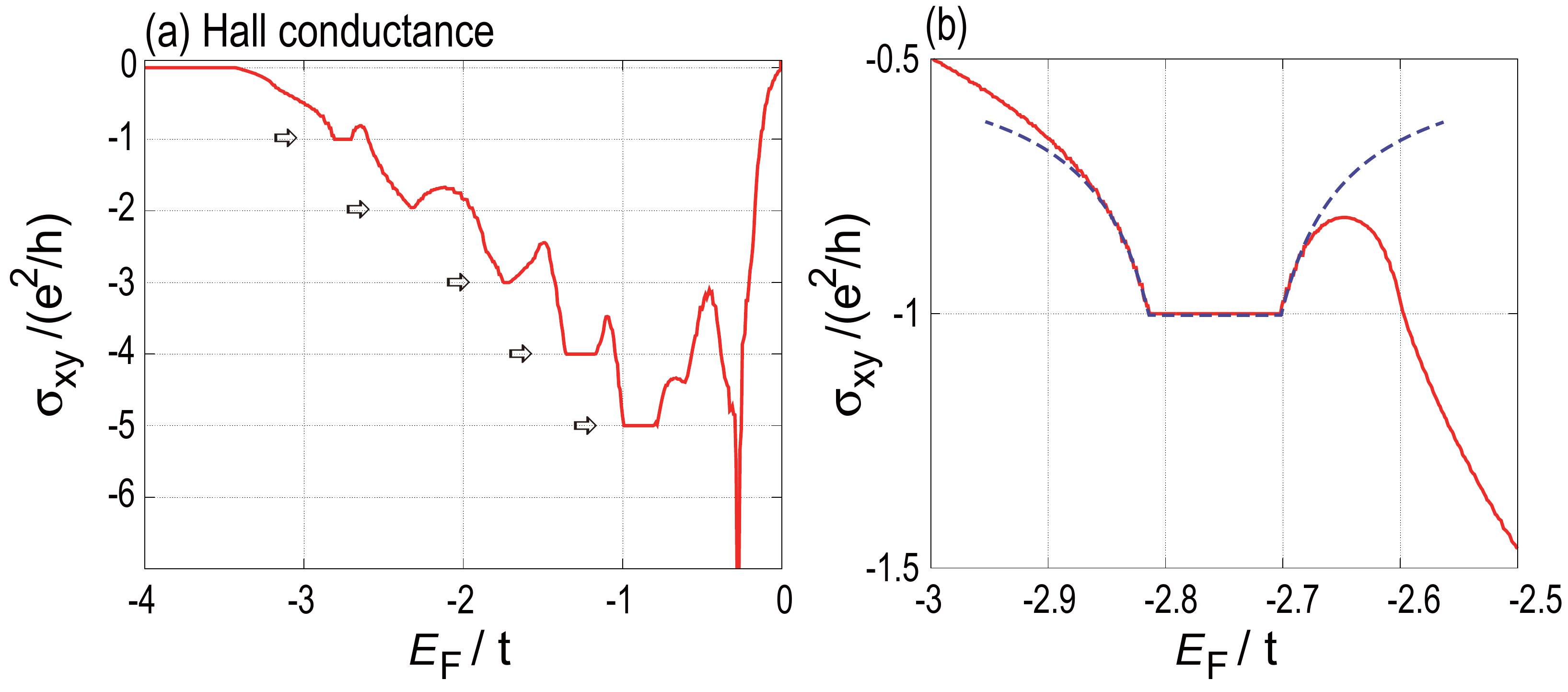}}
\caption{(a) The Hall conductance as a function of the chemical potential. 
The horizontal axis is the chemical potential, while the vertical axis is the Hall 
conductance $\sigma_{xy}$.
The Hall conductance is quantized as marked by arrows. 
There is a peculiar structure of dips in the conductance at the values where it is quantized.
(b) Zoom up of (a) near the band edge.
It is well fitted by the dotted curve obtained in the Dirac theory.}
\label{FigHall}
\end{figure}

\textbf{Hall conductance:}
The conductance is calculated by the Kubo formula\cite{TKNN,Kohmoto},
\begin{align}
\sigma_{xy}=&-\frac{ie^2}{h}\frac{2\pi}{L^2}\sum_{n,\mathbf{k}} f(E_{n\mathbf{k}}) \notag \\
&\quad\times\sum_{m(\neq n)}\frac{\langle n\mathbf{k} |\frac{\partial H}{\partial k_x}| m\mathbf{k}\rangle\langle m\mathbf{k} |\frac{\partial H}{\partial k_y}| n\mathbf{k}\rangle -(n\leftrightarrow m)}{(E_{n\mathbf{k}}-E_{m\mathbf{k}})^2}
\label{Kubo}
\end{align}
where $n$ and $m$ are the band indices and $f(x)$ is the Fermi distribution function.
We show the Hall conductance in Fig.\ref{FigHall} calculated using the Kubo formula (\ref{Kubo}).
At the zero temperature, it is quantized to be the Chern number when the Fermi energy is 
inside the gap,
$\sigma_{xy}=\frac{e^2}{h}\sum_{n:\text{filled}}C_n$,
where $C_n$ is the Chern number of the $n$th band.
Below the band gap, the Hall conductance decreases, while it increases above the band gap.
This is due to the fact that the sign of the Berry curvature is opposite between the two adjacent bands.
The total Chern number can be very large for large $\lambda$,
which is distinct from the QAHE in magnetic topological insulators~\cite{QAHE}. 

This behavior of the Hall 
conductance across the gap can be interpreted by the Dirac theory, 
which gives 
\begin{equation}
\sigma _{xy}=\left\{ 
\begin{array}{ccc}
-1/2 & \text{for} & \left\vert \mu \right\vert <\left\vert m\right\vert  \\ 
-m/(2\left\vert \mu \right\vert ) & \text{for} & \left\vert \mu \right\vert
>\left\vert m\right\vert 
\end{array}.
\right. \label{HallCon}
\end{equation}
It describes the peculiar behavior of the Hall conductance quite well as in Fig\ref{FigHall}(b), 
where the conductance is quantized inside the gap and continuously changes 
outside the gap showing a dip structure.

\begin{figure}[!t]
\centerline{\includegraphics[width=0.5\textwidth]{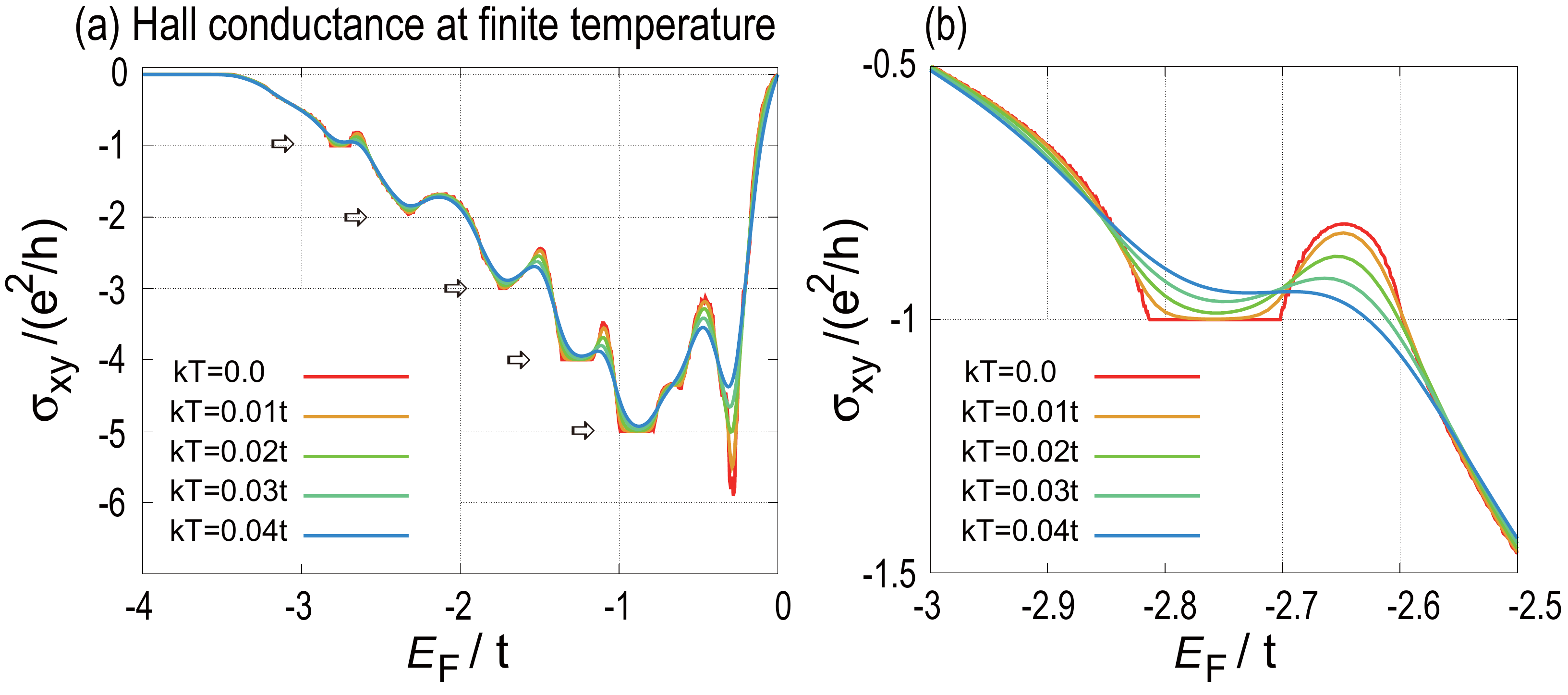}}
\caption{(a) The Hall conductance at finite temperature.
Though it is not quantized, the peculiar dip structure is observable even at room temperature.
(b) Zoom up of (a) near the band edge.
}
\label{FigFinite}
\end{figure}

\textbf{Finite temperature:}
We show the Hall conductance at finite temperature in Fig.\ref{FigFinite}.
It is evident that the temperature scale is given by the gap $\Delta$ in Fig. 1(b),
which is around $0.1t$ for $\lambda=2$.
Although the quantization of plateau is broken at 
$k_BT=0.01t\simeq 0.01eV\simeq 100K$,
the peculiar dip structure toward the formation of plateau is clearly 
visible even at $k_BT\simeq 400K$. 
Consequently, our prediction of QHE without external magnetic field can be 
experimentally observable at room temperature.

\begin{figure}[!t]
\centerline{\includegraphics[width=0.5\textwidth]{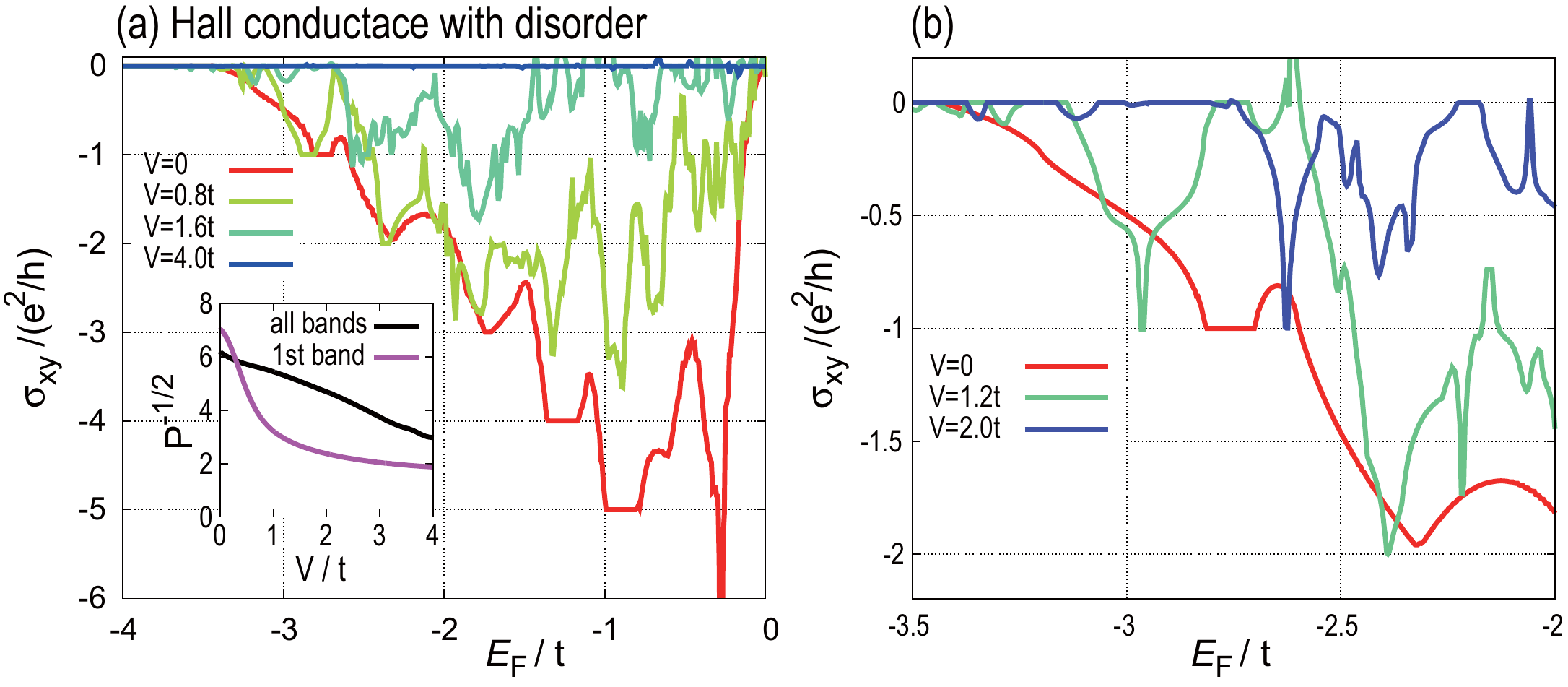}}
\caption{(a) The Hall conductance as a function of the chemical potential for various disorder strength. 
The horizontal axis is the energy, while the vertical axis is the Hall conductance $\sigma_{xy}$.
Disorder strength is indicated by color. Hall plateaux disappear as disorder increases.
However, the Hall plateau at $\sigma_{xy}=-e^2/h$ is transformed to the Hall plateau 
at $\sigma_{xy}=0$. 
(b) The Hall conductance in the vicinity of the band edge.
}
\label{FigHallimp}
\end{figure}

\textbf{Disorder effect:}
Up to now, we focus on the pure system, and the Berry curvature in momentum space.
Now we introduce the disorder potential as given by  
\begin{equation}
H_{\text{imp}}=\sum_i U_id^{\dagger}_id_i, 
\end{equation}
where $U_i$ takes a uniform random distribution for $-V<U_i<V$. 
We have calculated the Hall conductivity $\sigma_{xy}$
for the system of size $8 \times 8$, including 4 unit cells.
Figure 6 shows the numerical results, which clearly shows the 
plateau transition occurs at $V \cong $t for the lowest band, i.e.,
$\sigma_{xy}= - \frac{e^2}{h}$. Note that
when $|\sigma| > \frac{1}{2} \cdot \frac{e^2}{h}$, 
the two-parameter scaling trajectory converges to the quantized Hall state.
Taking this criterion, the lowest occupied band, which is most 
promising candidate to observe the QHE in SkX, can support 
the plateau up to $V \cong $t.  This magnitude of the random 
potential is similar to the separation between LLs 
in Fig. 1(c) and is much larger than the gap $\Delta$ in Fig. 1(b) 
which is $\cong 0.1$t in the present case. 
This is because the dispersion gives the stability or robustness 
of the extended state. 
For higher energy bands, the Hall conductivity is reduced 
more slowly as $V$ increases, and the criterion 
$|\sigma| > \frac{1}{2} \cdot \frac{e^2}{h}$
is satisfied up to $V \cong 3$t as shown in Fig.6(a).
We also calculated quantity $P_n$ for the eigenfunction 
$\langle i | \psi_n\rangle $ with energy $\varepsilon_n$ defined  by 
\begin{equation}
P_n = \sum_i |\langle i | \psi_n\rangle |^4,
\end{equation}
which measures the extent of the wavefunction. 
Namely, when $\langle i | \psi_n\rangle$ extends over the $M$ sites,
$P_n \sim M^{-1}$. Therefore, the localization length 
is estimated by $P_n^{-1/2}$ in 2D. 
We show in the inset of Fig. 6 the averaged $P_n^{-1/2}$
for the lowest band (purple curve) and all the bands (black curve) as 
a function of $V$.  It seen that the disappearance of the QHE roughly corresponds 
to the localization of the wavefunctions less than 4, where   
4 is the size of the unit cell (skyrmion). 
Therefore, since the mean free path $\ell$ is always longer than the localization length,
the QAHE is observed only in the case where $\ell > 2 \lambda$, i.e., the Berry   
curvature in momentum space is more appropriate picture rather than 
that in the real space.

\textbf{Discussions:}
We have shown that the QAHE occurs in metallic SkX 
due to the emergent magnetic field induced by the spin texture.
It is shown that there is a gap between the lowest and second-lowest bands. 
The electron density should be of the order of one electron per one skyrmion, 
whose electron density is given by $1/4\lambda^2$. In order to realize experimental 
situations, it is necessary to reduce the number of electrons. In this sense, 
an interface of dilute magnetic semiconductors will be promising.
The emergent magnetic field and the associated energy gap is proportional 
to $1/\lambda^2$. Namely, skyrmions with smaller 
radius are better for realizing QAHE.

The QAHE is robust against disorder as in the case of system with corresponding 
uniform magnetic field with Landau levels. 
Considering the very large mean magnetic field of
$\sim 4000$T for the skrmion size $\lambda \sim 1$nm,
the disorder does not destroy the QAHE so seriously. This is because
the change in the Chern numbers occurs only via the pair annihilation,
and the appreciable energy dispersion of the band in SkX protects the
Chern number. 
On the other hand, the band gap $\Delta$ of SkX is smaller than $\Delta_0$, i.e.,
that of the corresponding system with uniform magnetic field, and hence the stability
against the thermal excitation is reduced.  
$\Delta$ is estimated as $\Delta_0/5$, but as mentioned above,
the emergent magnetic flux induced by SkX is gigantic 
and even the QAHE even at room temperature is expected. 

This work was supported in
part by Grants-in-Aid for Scientific Research from the Ministry of
Education, Science, Sports and Culture No. 24224009 and 25400317.


\begin{thebibliography}{99}
\bibitem{SkRev} N. Nagaosa and Y. Tokura, Nat. Nanotech. \textbf{8} 899 (2013)

\bibitem{Bogdanov}
N. Bogdanov and D.A. Yablonskii, Sov. Phys. JETP
68, 101 (1989).

\bibitem{BogdanovB}
N. Bogdanov and A. Hubert, J. Magn. Magn. Mater. 138, 255 (1994).

\bibitem{BogdanovC}
U.K. Roessler,  N. Bogdanov  and C. Pfleiderer, Nature 442, 797 (2006).

\bibitem{Mol} S. Mhlbauer et. al., Science
323, 915 (2009)

\bibitem{Yu} X.Z. Yu et. al., Nature 465, 901 (2010).

\bibitem{Heinze} S. Heinze et. al., Nature Phys. 7, 713 (2011).

\bibitem{Shulz} T. Schulz, R. Ritz, A. Bauer, M. Halder, M. Wagner, C. Franz, C. Pfleiderer, K. Everschor, M. Garst and A. Rosch, Nature Physics  8, 301 (2012)

\bibitem{Lee} M. Lee, W. Kang, Y. Onose, Y. Tokura and N.P. Ong, Phys. Rev. Lett. 102, 186601 (2009).

\bibitem{Li} Li. Yufan, N. Kanazawa, X. Z. Yu, A. Tsukazaki, M. Kawasaki, M. Ichikawa, X. F. Jin, F. Kagawa, and Y. Tokura, Phys. Rev. Lett. 110, 117202 (2013).

\bibitem{Kanazawa} N. Kanazawa et. al., Phys. Rev. Lett. 106, 156603 (2011).

\bibitem{Neu} Neubauer, A. et. al. Phys. Rev. Lett. 102, 186602 (2009).

\bibitem{FeGe} X. Z. Yu, N. Kanazawa, Y. Onose, K. Kimoto, W. Z. Zhang, S. Ishiwata, Y. Matsui and Y. Tokura, Nature Materials  10, 106 (2011)

\bibitem{Taguchi} Y. Taguchi, Y. Oohara, H. Yoshizawa, N. Nagaosa, and Y. Tokura, Science \textbf{291}, 2573 (2001).

\bibitem{Onoda} M. Onoda, G. Tatara, N. Nagaosa, J. Phys. Soc. Jpn. 73, 2624 (2004)

\bibitem{QAHE} C.Z. Chang, et al. Science 340, 167-170 (2013).

\bibitem{Anderson} P. W. Anderson and H. Hasegawa, Phys. Rev. 100, 675 (1955).

\bibitem{Ohgushi} K. Ohgushi, S. Murakami and N. Nagaosa,  Phys. Rev. B \textbf{62} 6065(R) (2000)

\bibitem{Raja} R. Rajaraman, Solitons and Instantons (Elsevier, 1987)

\bibitem{Volovik} G. Volovik, Journal of Physics C: Solid State Physics 20, L83 (1987).

\bibitem{Zhang} S. Zhang and S. S.-L. Zhang, Phys. Rev. Lett., 102, 086601 (2009).

\bibitem{Zang} J. Zang, M. Mostovoy, J. H. Han, and N. Nagaosa, Phys. Rev. Lett. 107, 136804 (2011)

\bibitem{TKNN} D. J. Thouless, M. Kohmoto, M. P. Nightingale and M. den Nijs, Phys. Rev. Lett. 49, 405 (1982)

\bibitem{Kohmoto} M. Kohmoto, Ann. Phys. (N.Y.) 160, 343 (1985)
\end{thebibliography}
\end{document}